\documentclass[amsmath, amssymb,10pt, aps, prb, twocolumn, notitlepage, longbibliography, superscriptaddress]{revtex4-1}

\usepackage{graphicx}
\usepackage{calc}
\usepackage{float}
\usepackage{braket}
\usepackage{slashed}
\usepackage{wasysym}
\usepackage{dsfont}
\usepackage{amsthm,amsmath,amsfonts,amssymb,verbatim,color}
\usepackage{lipsum}
\usepackage{bm}
\usepackage{epsfig,slashed}
\usepackage{hyperref}
\usepackage{flafter}
\usepackage{booktabs}
\usepackage{makecell} 
\usepackage{bbm} 
\usepackage[export]{adjustbox} 

\newcommand{\basictetrahedron}{\raisebox{1ex}{\rotatebox{200}{\scalebox{1}[0.3]{$\bigtriangledown$}}}%
	\hspace*{-0.86em}\raisebox{-0.2ex}{\rotatebox{80}{\scalebox{1}[0.3]{$\bigtriangledown$}}}%
	\hspace*{-0.42em}\raisebox{1.58ex}{\rotatebox{320}{\scalebox{1}[0.3]{$\bigtriangledown$}}}
}

\newcommand{\tetrahedron}{\ifmmode%
	\raisebox{-0.35ex}{\basictetrahedron}%
	\else%
	\basictetrahedron%
	\fi}

\newcommand{\bk}{{\bf k}}

\newcommand{\br}{{\bf r}}
\newcommand{\bs}{{\bf s}}

\def\brho{{\boldsymbol \rho}}

\newcommand{\cO}{{\cal O}}

\begin{document}

\title{Effect of vacancy defects on geometrically frustrated magnets}

\author{Sergey Syzranov}
\affiliation{Physics Department, University of California, Santa Cruz, California 95064, USA}

\begin{abstract}
Quenched disorder may prevent the formation of the widely sought quantum-spin-liquid states (QSLs) 
or mask their signatures by inducing a spin-glass state, which is why considerable experimental efforts are directed at purifying 
materials that may host QSLs. 
However, in geometrically frustrated (GF) magnets, the largest class of materials in which QSLs are sought,
the glass-transition temperature $T_g$ grows with decreasing the density of vacancy defects, accompanied by a simultaneous 
growth of the magnetic susceptibility.
In this paper, we develop a phenomenological theory of glass transitions and magnetic susceptibility 
in 3D geometrically frustrated (GF) magnetic materials.
We consider a model of a GF magnet in which the glass transition occurs in the absence of vacancies, e.g., due to other types of quenched disorder.
We show that disorder that creates weak local perturbations, e.g. weak random strain, leads to the growth of the transition temperature $T_g$.
By contrast, vacancies reduce $T_g$ for small vacancy concentrations.
Another consequence of the presence of vacancies is the creation of quasispins,
effective magnetic moments localised near the vacancies, that contribute to the magnetic susceptibility of the system together with the bulk spins.
We show that increasing the vacancy density leads to an increase of the total magnetic susceptibility.
\end{abstract}


\maketitle


Quenched disorder (impurities, irregularities, vacancies, etc.) is one of the main obstacles 
to observing widely-sought quantum-spin-liquid (QSL) states~\cite{SavaryBalents:review}. Quenched disorder may not only hide the 
QSL signatures, such as the temperature dependence of the heat capacity, but, in three-dimensional (3D) systems~\cite{McMillan:firstNoTwoDGlass,HartmannYoung:TwoDGlass,Carter:TwoDGlass,Amoruso:TwoDGlass,Fernandez:TwoDGlass,Rieger:TwoDGlass}, also
induce the spin-glass state that may be incompatible with a QSL.
At present, the effect of quenched disorder on QSLs and magnetic materials 
expected to support QSL states is far from being understood.

\begin{figure}[H]
	\centering
	\includegraphics[width=\linewidth]{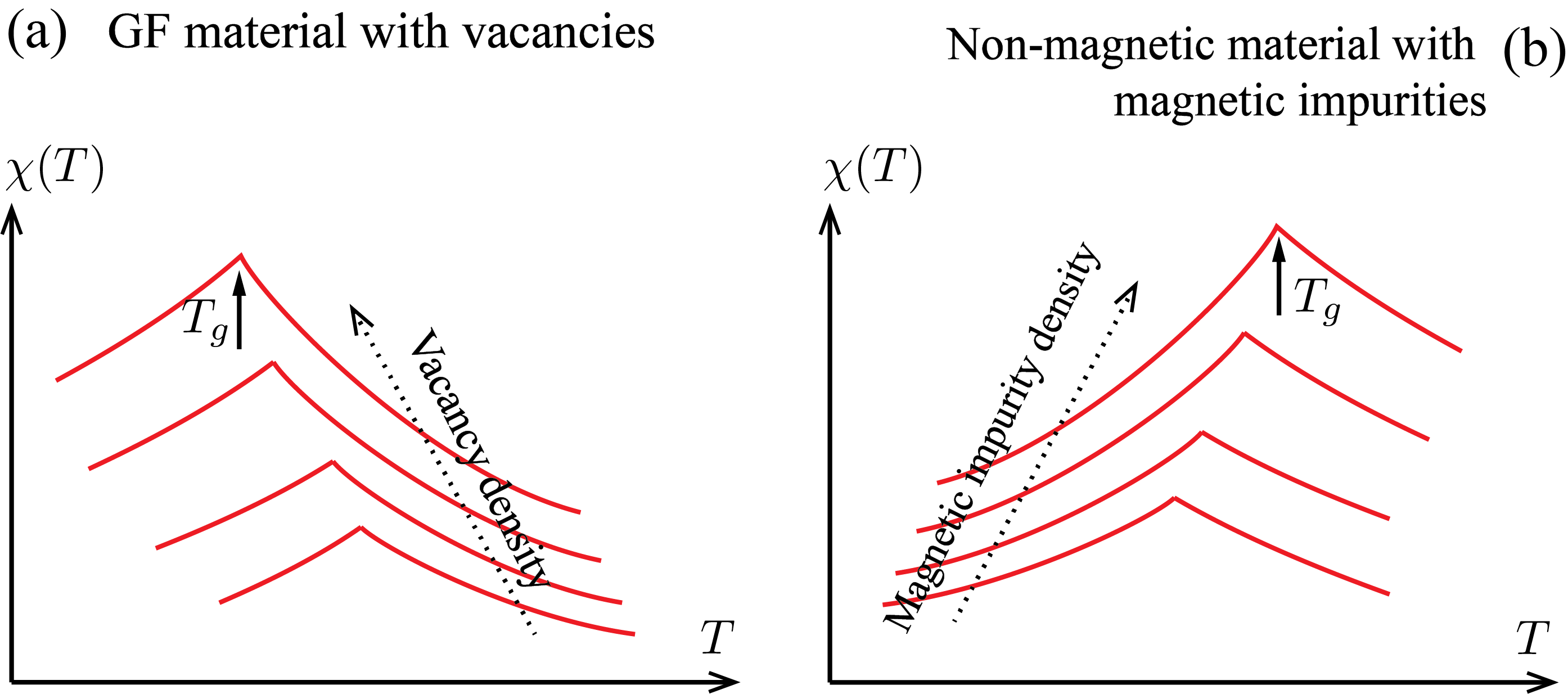}
	\caption{	\label{fig:susceptibilitytemperaturesummary} Schematic of the behaviour~\cite{Syzranov:HiddenEnergy} of the
		susceptibility $\chi(T)$ and glass-transition temperature $T_g$ in 
		(a) geometrically frustrated magnets and (b) conventional spin glasses for various densities of defects.
	}
\end{figure}

A common expectation, confirmed by the existing theories of spin glasses~\cite{MezardParisiVirasoro:book,Dotsenko:GlassReview,BinderYoung:review,SaundersChalker:pyrochloreMCHeisenberg,Andreanov:MCpyrochlore,Syzranov:HiddenEnergy}, is that purifying a material
makes random spin freezing less favourable and thus lowers the glass-transition temperature,
which, in turn, should make QSL signatures more vivid.
Available experimental data, however, for spin-glass transitions in geometrically frustrated (GF) magnets~\cite{Ramirez:FrustrationReview}, the largest
class of materials in which QSLs are sought, reveals several surprising trends~\cite{Syzranov:HiddenEnergy},
summarised in Fig.~\ref{fig:susceptibilitytemperaturesummary}, that contradict common intuition.

The most common type of quenched disorder in GF materials is vacancies, i.e. randomly located nonmagnetic atoms 
that replace magnetic atoms of the GF medium.
With {\it decreasing} the concentration of vacancies in GF materials, which contribute to quenched disorder but of which they may not be the only source,
the glass-transition temperature {\it increases}, reaching a finite value $T^*$, the ``hidden energy scale'',
in the limit of a vacancy-free material~\cite{Syzranov:HiddenEnergy}.
While vacancies in many GF materials are the only known source of disorder, such a spin freezing may present a challenge for
the observation of QSLs.

The origin of the hidden energy scale still remains to be explored. In most frustrated materials, it has the same order
of magnitude, $T^*\sim 10K$, significantly exceeded by the exchange couplings between the spins. In some materials, exemplified
by $Y_2 Mo_2 O_7$ \onlinecite{Thygesen:YMoO}, the hidden energy scale may be presumed to come from sources of disorder other than vacancies, such as random strain fields due to
the dynamic Jahn-Teller effect. In other materials, however, there
are no other significant known sources of disorder distinct from vacancies.
The origin and the value of the hidden energy scale, thus, require a careful further investigation.

Another remarkable universal trend that GF systems exhibit and that 
persists regardless of the source of disorder is the growth of the 
magnetic susceptibility $\chi(T_g)$
with decreasing the critical temperature $T_g$, $\frac{d\chi(T_g)}{dT_g}<0$
(see Fig.~\ref{fig:susceptibilitytemperaturesummary}a), when changing 
the concentration of vacancies. This trend holds for {\it all} GF magnets, for whom experimental
data on spin glasses are available~\cite{Syzranov:HiddenEnergy,RamirezCooper:SCGO,Greedan:YMoO,Ying:YMoO,Martinez:SCGO,Nambu:NiGaS,Oseroff:CdMnTe,Ratcliff:ZnCrO,Okamoto:NaIrO,Balodhi:NaIrO,LaBarre:FeTiO},
and is in contrast with the opposite trend, $\frac{d\chi(T_g)}{dT_g}>0$, in conventional spin glasses (Fig.~\ref{fig:susceptibilitytemperaturesummary}b).

The growth of the susceptibility with vacancy concentration is consistent with the empirical picture
of ``orphan'' spins or quasispins~\cite{Schiffer:TwoPopulationF,LaForge:QuasispinsZnCrGaO,SenMoessner:FractionalSpin,WollnyVojta:FractionalQuasispin}:
the shielding of a vacancy by the spins creates a degree of freedom that acts as 
a magnetic moment (a ``quasispin''). As a result, adding vacancies to a GF material leads to the growth of the susceptibility $\chi(T)$
(Fig.~\ref{fig:susceptibilitytemperaturesummary}a), similarly to the growth of susceptibility 
observed
in a non-magnetic medium (conventional spin glass) when adding magnetic impurities (Fig.~\ref{fig:susceptibilitytemperaturesummary}b).
Empirically, the contribution of the quasispin degrees of freedom may be separated from that of the {\it bulk spins},
i.e. spins sufficiently far from the vacancies, which behave as an independent subsystem 
(e.g. have a different Weiss constant~\cite{Schiffer:TwoPopulationF,LaForge:QuasispinsZnCrGaO}).

Quasispins alone cannot account for the observed~\cite{Syzranov:HiddenEnergy} decrease of the glass transition 
temperature with increasing the concentration of vacancies; in a system of randomly located magnetic moments,
the glass transition temperature grows with the density of the moments~\cite{MezardParisiVirasoro:book,Dotsenko:GlassReview,BinderYoung:review,SaundersChalker:pyrochloreMCHeisenberg,Andreanov:MCpyrochlore,Syzranov:HiddenEnergy}.
If one assumes that the bulk spins in the GF medium undergo freezing at the transition,
instead of the vacancy-induced quasispins, then
diluting such bulk spins by vacancies may be qualitatively consistent with lowering the glass transition temperature. However, diluting the bulk spins by vacancies
also decreases their contribution to the magnetic susceptibility.
It is natural to assume, therefore, that describing the behaviour of both the susceptibility $\chi(T)$ and the glass transition temperature $T_g$
should involve both the quasispin degrees of freedom and the degrees of freedom of the bulk spins.

In this paper, we develop a phenomenological theory of the effect of vacancy defects on the properties 
of frustrated magnetic materials. We provide a scenario that explains simultaneously the observed, previously unexplained
trends: the growth of the
glass transition temperature with decreasing vacancy concentration, the growth of the magnetic susceptibility with vacancy concentration
and the $d\chi(T_g)/dT_g<0$ trend observed in GF magnets. It also predicts the growth of the glass-transition temperature with
increasing the strength of disorder that creates weak local perturbations in the system, such as weak random strain.

{\it Qualitative picture.}
We assume that the material undergoes 
a spin-glass phase transition in the absence of vacancies. Such a
transition may be driven by quenched disorder other than randomly located vacancies, such as 
random strain fields in $Y_2 Mo_2 O_7$ \onlinecite{Thygesen:YMoO}, 
 or conceivably occur even in the absence of disorder~\cite{Villain:GlassWithoutDisorder},
 similarly to the conventional liquid-glass transition~\cite{Parisi:GlassBook}.
The exact mechanism of the glass transition in a vacancy-free system is not important for our consideration. We assume that averaging of observables over the positions of vacancies and
realisations of other forms of disorder can be carried out independently.

The temperature $T_g$ of the glass
 transition is assumed to be significantly exceeded by the absolute value of the 
Weiss constant $|\theta_W|$, which is on the order of the energies of 
isolated spin-flip excitations in the GF medium [for instance, for isotropic exchange interactions with the 
coupling $J$, the mean field approximation gives 
$\theta_W=-\frac{1}{3}ZJs(s+1)$, where $Z$ is the coordination number and $s$ is the value of the spin].
As a result, isolated spin-flip excitations are strongly suppressed near the glass transition.

We show that
at temperatures $T_g\lesssim T\ll |\theta_W|$, vacancies create, by breaking some of the bonds
in the GF medium, degenerate states of the system with different magnetizations, which is equivalent to the existence of magnetic degrees of freedom
(quasispins) associated with the vacancies.
In a broad temperature interval, the magnetic susceptibility may be expected to obey the formula
\begin{align}
	\chi(T)=\frac{A(n-n_\text{imp})}{T+|\theta_W|}
	+\frac{Bn_\text{imp}}{T},
	\label{GenericSusceptibility}
\end{align}
which interpolates between the Curie-Weiss contribution of the bulk spins at $T\gtrsim |\theta_W|$ and the 
contribution of the vacancies at $T\ll |\theta_W|$; $n_\text{imp}$ is the concentration of the vacancies.

At the same time, the dilution of the bulk-spin medium by vacancies reduces
correlations between the bulk spins and lowers the glass transition temperature, as we show below.
The glass transition is thus driven by the degrees of freedom of the bulk spins 
and not by the quasispins associated with the vacancies.

{\it Glass transition temperature and vacancies.} 
The transition can be detected via the glass order parameter 
$Q_\br^{\alpha\beta}=\left<\hat \bs_\br^\alpha\cdot 
\hat \bs_{\br}^\beta\right>$, 
the correlator of spins in different replica subspaces~\cite{MezardParisiVirasoro:book,BinderYoung:review} $\alpha$
and $\beta$, where $\langle\ldots \rangle$ is averaging over the states of the spin and non-vacancy
disorder, if present, for particular locations of the vacancies.
The order parameter $Q_\br^{\alpha\beta}$ is finite 
at temperatures below the glass-transition temperature, $T<T_g$,
and vanishes above the transition, at $T>T_g$.

The decrease of the transition temperature $T_g$ when adding vacancies can be 
understood intuitively as a result of the decrease of the average coupling between the bulk spins of the GF medium that undergo
the glass freezing.
To describe this effect quantitatively, we assume that in the absence of vacancies the glass transition can be described by a mean-field replicated 
free energy
\begin{align}
	F(Q)=\frac{1}{2}Q_\br^{\alpha\beta} \left(K^{-1}\right)_{\br\br^\prime} Q_{\br^\prime}^{\alpha\beta}+
	\cO\left( Q^3\right),
	\label{FreeEnergyClean}
\end{align}
in which the lowest eigenvalue of the
matrix $\left(K^{-1}\right)_{\br\br^\prime}$
vanishes when approaching the transition:
\begin{align}
	K_{\bk=0}^{-1}=a(T_g-T),
	\label{KonTgDependence}
\end{align}
where $K_\bk$ is the Fourier-transform of the correlator $K_{\br\br^\prime}$; summation over repeated indices is implied
and $a>0$.
The free energy \eqref{FreeEnergyClean} describes the action of
a vacancy-free system, averaged over the realisations of the other types of disorder, if present.
Therefore, 
the matrix $K_{\br\br^\prime}$ is translationally invariant.

For simplicity, we consider a system in which $Q_\br^{\alpha\beta}$ does not have time dynamics, corresponding, e.g., to 
models with effectively classical spins. However, the argument developed here applies as well in the presence of quantum dynamics of spins, i.e.
for a system described by the action \eqref{FreeEnergyClean} with the fields $Q$ that depend on the Matsubara time $\tau$ in addition
to the coordinates.

We note that for a particular pair of replicas $\alpha\neq\beta$, the corresponding quadratic form in
Eq.~\eqref{FreeEnergyClean} is negative above the transition, at $T>T_g$.
However, the number $n(n-1)/2$ of such pairs is also negative in the replica limit $n\rightarrow 0$
[for example, for a replica-symmetric order parameter $Q_\br^{\alpha\beta}=q$, the free energy is given by 
$\tilde{F}=\lim\limits_{n\rightarrow 0} \frac{F(q)}{n}\propto a(T-T_g)$], resulting in a positive action for $T>T_g$.

Because the free energy \eqref{FreeEnergyClean} is quadratic in small $Q_\br^{\alpha\beta}$,
the fluctuations of the order parameter $Q_\br^{\alpha\beta}$ may be considered Gaussian 
above the transition ($T>T_g$). This allows one to describe the effects of vacancy defects and other small perturbations on
the glass transition diagrammatically, with the matrix $K_{\br\br^\prime}$ playing the role of the Green's
function.

To consider the effect of a single vacancy or other defect added to the system,
we add the term $\lambda \left(Q_\brho^{\alpha\beta}\right)^2$ to the action \eqref{FreeEnergyClean}
at site $\brho$.
For $\lambda\rightarrow -\infty$, such a term leads to the vanishing of the field $Q^{\alpha\beta}_\brho$
and mimics a vacancy at site $\brho$. For small $\lambda>0$, a combination of such terms can mimic, for example, the effect of
weak fluctuations of the exchange coupling averaged over that coupling~\cite{Andreanov:MCpyrochlore,Syzranov:HiddenEnergy}.
The presence of one defect modifies the correlator $K_{\br\br^\prime}=\langle Q_{\br}^{\alpha\beta}Q_{\br^\prime}^{\alpha\beta}\rangle/T$
of the fields $Q^{\alpha\beta}$ (see Supplemental Material~\cite{SupplementalMaterial}):
\begin{align}
	K_{\br\br^\prime}\rightarrow
	K_{\br\br^\prime}-K_{\br\brho}\frac{\lambda}{1+\lambda K_{\brho\brho}}K_{\brho\br^\prime}.
	\label{CorrOneDef}
\end{align}
We note that the quantity $K_{\brho\brho}=\int\frac{d\bk}{(2\pi)^d} K_\bk$ is negative above the transition ($T>T_g$), non-singular in $T-T_g$
for quadratically dispersive modes [$K_\bk^{-1} \propto a(T_g-T)-b\bk^2$] in 3D and independent 
of the choice of the site $\brho$. 
For $\lambda\rightarrow -\infty$, corresponding to a vacancy defect, the correlator~\eqref{CorrOneDef} in the presence of the vacancy
vanishes for $\br=\brho$ or $\br^\prime=\brho$ because $Q_\brho^{\alpha\beta}=0$ at the location of the vacancy.

In the limit of dilute defects, the disorder-averaged correlator $\tilde{K}$
of the fields $Q_\brho^{\alpha\beta}$ is given by
\begin{align}
	\tilde{K}_\bk^{-1}=K_\bk^{-1}+\frac{\lambda n_\text{imp}}{1+\lambda K_{\brho\brho}},
\end{align}
where $n_\text{imp}$ is the defect density and we have used that the contributions of dilute defects to the ``self-energy'' 
$\Sigma_\bk=K_\bk^{-1}-\tilde{K}_\bk^{-1}$ are additive.
The effect of the impurities is, therefore, equivalent to the shift 
of temperature
\begin{align}
	T_g \rightarrow T_g +\frac{n_\text{imp}\lambda}{a\left(1+\lambda K_{\brho\brho}\right)}.
\end{align}
For $\lambda\rightarrow -\infty$, corresponding to vacancy defects, the glass transition temperature is lowered
by $n_\text{imp}/\left(a\left|K_{\brho\brho}\right|\right)$, in accordance with
experimentally observed trends~\cite{Syzranov:HiddenEnergy}. By contrast, quenched disorder that creates weak local perturbations 
($0<\lambda<|K_{\brho\brho}|^{-1}$) increases $T_g$.

{\it Magnetic susceptibility of vacancies.}
In what immediately follows, we demonstrate microscopically the emergence of the quasispin degrees of freedom in a frustrated medium.
We assume first that sources of quenched disorder distinct from vacancies are absent. If a disorder-free glass transition is possible, we also assume that
the temperature of the system exceeds the transition temperature $T_g$. 
We consider a GF material whose ground states have zero magnetization. 
%
At a finite temperature, the fluctuations of the magnetization in the absence of vacancies lead to a finite magnetic Curie-Weiss 
susceptibility given by the first term in Eq.~\ref{GenericSusceptibility}.

The presence of vacancies removes some of the bonds in the GF lattice and allows, in general, for degenerate
ground states with nonzero magnetizations. Degenerate states with different magnetizations give large Curie-type
contributions $\chi(T)\propto n_\text{imp}/T$ to
the magnetic susceptibility equivalent to that of free spins at low temperatures $T_g<T\ll |\theta_W|$.

To illustrate the emergence of such quasispin degrees of freedom, we consider Ising spins $s_\br=\pm 1$ on a pyrochlore lattice, a frustrated lattice
consisting of tetrahedra touching at the corners. In the absence of vacancies, the Hamiltonian of this model is given by (see, e.g., Refs.~\onlinecite{IsakovSondhi:firstCoulomb,Henley:secondCoulomb,Henley:review})
\begin{align}
	H=J\sum_{(\br\br^\prime)} s_\br s_{\br^\prime} 
	=\frac{J}{2}\sum_i S_i^2+ \text{const},
	\label{CleanIsingHamiltonian}
\end{align}
where the summation in the first sum is carried out over all nearest-neighbour
pairs of spins $\br\br^\prime$; $J>0$ is the antiferromagnetic coupling constant; the last expression represents the sum
over all the tetrahedra $i$ that constitute the pyrochlore lattice and 
$S_i=\sum\limits_{\br\in\basictetrahedron_i} s_\br$ is the sum of the spins of the $i$-th tetrahedron.

In the absence of vacancies, the ground states correspond to the vanishing spins of all tetrahedra, $S_i=0$, 
with two spins pointing up and two spins pointing down in each tetrahedron~\cite{IsakovSondhi:firstCoulomb,Henley:secondCoulomb,Henley:review} (``two-in, two-out'' rule). 
Such states have a degeneracy that scales exponentially with the system size and can be parametrised 
by a Coulomb field (``Coulomb phase''~\cite{IsakovSondhi:firstCoulomb,Henley:secondCoulomb,Henley:review}).

A vacancy at location $\brho$ can be modeled by adding the Hamiltonian $H_\brho=-J\sum_{(\brho\br)}s_\brho s_\br $
to the Hamiltonian \eqref{CleanIsingHamiltonian}, which results in the effective cancellation of the exchange couplings next to 
the vacant site.
The Hamiltonian of the system with such a vacancy at location $\brho$ is given, up to an additive constant, by
\begin{align}
	\tilde{H}=\frac{J}{2}\left(S_1-s_\brho\right)^2+\frac{J}{2}\left(S_2-s_\brho\right)^2
	+\frac{J}{2}\sum_{i\neq 1,2} S_i^2,
	\label{EnergyIsing}
\end{align}
where $S_1$ and $S_2$ are the total spins (including $s_\brho$) of the two tetrahedra that share site $\brho$.

\begin{figure}[H]
	\centering
	\includegraphics[width=\linewidth]{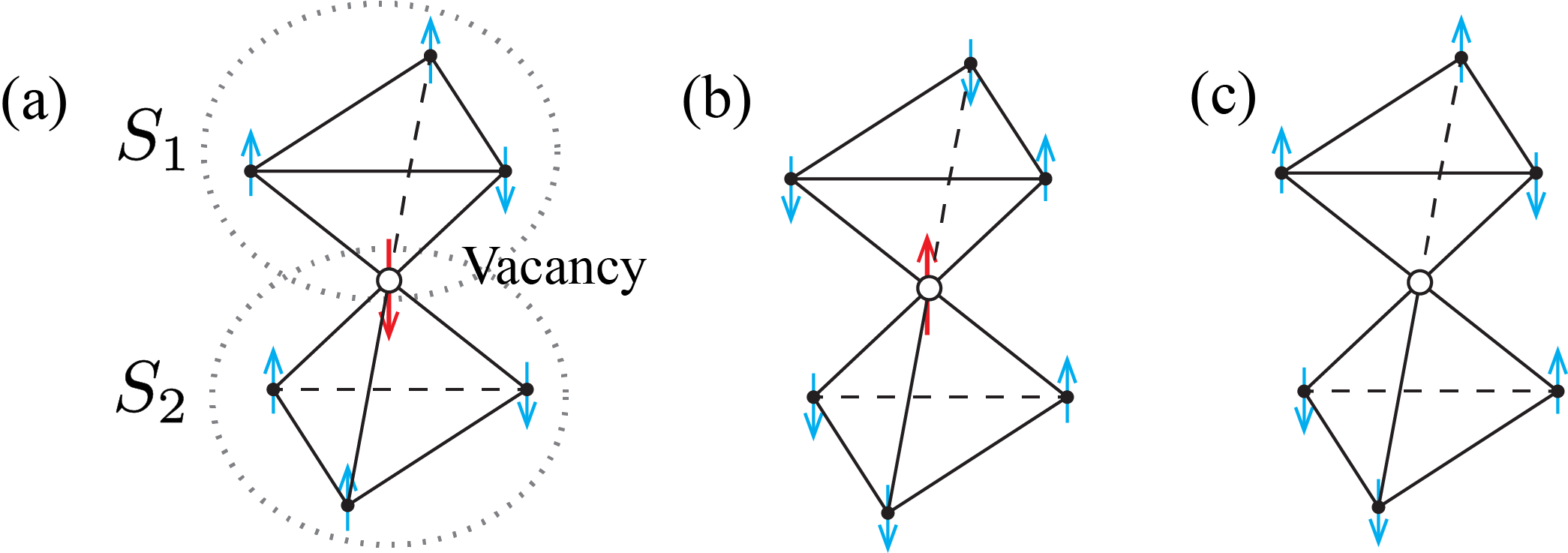}
	\caption{\label{fig:spinconfigurations} Spin configurations near a vacancy on a pyrochlore lattice. States (a) and (b) are obtained by removing a spin
		(shown in red) pointing, respectively, down and up from a ground state of a vacancy-free system.
	State (c) is obtained by removing a site from an excited state obtained from a ground state of a vacancy-free system by flipping 
	all spins in a semi-infinite chain that starts at the removed site. The three states have the same energy.}
\end{figure}

Because $|S_{1,2}-s_\brho|\geq 1$, the minimum energy of the system with a vacancy is given by $J$.
There are three types of such ground states, shown in Fig.~\ref{fig:spinconfigurations}: a) $S_1-s_\brho=S_2-s_\brho=1$, with two spins in the vacancy-sharing tetrahedra
pointing up and one spin pointing down, b) $S_1-s_\brho=S_2-s_\brho=-1$, with two spins in the vacancy-sharing tetrahedra
pointing down and one spin pointing up and c) $S_1-s_\brho=\pm 1$, $S_2-s_\brho=\mp 1$ (with $S_i=0$ for $i\neq 1,2$).

States (a) and (b) are obtained by removing a spin pointing, respectively, down and up from a ground state of a vacancy-free system. Because the latter has zero
total magnetization (spin), states (a) and (b) have total spins of $+1$ and $-1$,
respectively, and a degeneracy of $z/2$, where $z$ is the degeneracy of 
ground states in a vacancy-free system. The magnetization comes from tetrahedra $1$ and $2$
and is thus concentrated near the vacancy.

State (c) has zero magnetization and 
is obtained by removing a spin from an excited state of the vacancy-free system in which one tetrahedron 
has a magnetization of $2$ and the other tetrahedra have zero magnetization. In the representation of the Coulomb phase,
such an excited state corresponds to a monopole excitation at the centre of the respective tetrahedron. Each such excited state
is obtained from a ground state of the vacancy-free system by flipping the direction of the Coulomb field along a semi-infinite
chain (``Dirac string'')~\cite{IsakovSondhi:firstCoulomb,Henley:secondCoulomb,Henley:review}.
The degeneracy of states (c) is, therefore, on the order of or smaller than the
degeneracy of states (a) and (b). 

The existence of same-energy states with different finite magnetizations localised near the vacancy leads to a finite variance 
$\langle S^2\rangle\sim 1$ of the total spin and, according to the fluctuation-dissipation theorem, a finite magnetic susceptibility
$\chi_1(T) =\frac{g^2\mu_B^2 \langle S^2\rangle}{T}$ associated with one vacancy.
For sufficiently dilute vacancies, their contributions are additive, and the magnetic susceptibility exhibits the 
$\chi(T)\propto n_\text{imp}/T$ behaviour.

An alternative approach to deriving quasispin variables has been applied in Ref.~\onlinecite{SenMoessner:FractionalSpin}. In this approach,
a GF system is described by Gaussian magnetic fluctuations with phenomenologically introduced parameters. 
By integrating out those fluctuations on the $SrCr_{9x}Ga_{12-9x}O_{19}$ lattice with a vacancy, it has been found in Ref.~\onlinecite{SenMoessner:FractionalSpin}
that a vacancy is equivalent to a classical fractional spin in terms of its response to an external field. A fractional (quasi-)spin associated with a vacancy 
in a triangular lattice has also been demonstrated numerically in Ref.~\onlinecite{WollnyVojta:FractionalQuasispin} in the large-$S$ approximation.

We expect that similar quasispin degrees of freedom associated with vacancies emerge 
generically for all GF lattices that have zero average magnetization in ground states. The removal of a site in such lattices may be
expected to lead to nonzero magnetizations of the ground states of the lattice with a vacancy. In the presence of symmetry with respect to flipping 
or rotating all spins, the effect of a vacancy is then equivalent to the presence of a free magnetic moment in the system.
We leave, however, a microscopic investigation of such quasispins and their values for other specific frustrated lattices for future studies.

Introducing vacancies to the system increases the quasispin contribution to the magnetic susceptibility. 
At the same time, it reduces the density of the bulk spins, thus suppressing their contribution.
As can be seen from Eq.~\eqref{GenericSusceptibility}, however, the total magnetic susceptibility grows with increasing the density of vacancies, $d\chi(T)/dn_\text{imp}>0$,
in the dilute limit at low temperatures $T\ll|\theta_W|$ (assuming the constants $A$ and $B$ are of the same order of magnitude).

{\it Interplay of quasispins with other sources of quenched disorder.}
For the model we consider in this paper, the glass transition is driven by sources of disorder other than vacancies.
Non-vacancy quenched disorder may, in principle, be varied in experiment by substituting non-magnetic atoms in a GF compound by 
non-magnetic impurity atoms.
Non-vacancy quenched disorder lifts the degeneracy between the ground states we have discussed.
If such a lifting is sufficiently weak, its effect on the quasispins is equivalent to that of a random magnetic field.

The behaviour of the quasispin contribution to the susceptibility will be unaltered at temperatures $|\theta_W|\gg T\gg \Delta E$ significantly exceeding the
characteristic energy gap $\Delta E$ between the states of a quasispin.
The contribution of the vacancies to the susceptibility, however, significantly changes at low temperatures $T\ll \Delta E$ (possibly below the glass
transition). Due to random orientation of the effective magnetic field acting on the quasispins, 
the susceptibility of the vacancies is on the order of the transverse susceptibility of a two-level system with the splitting $\Delta E$,
which gives the estimate for the contribution of the vacancies to the magnetic susceptibility $\chi_\text{vac}\sim \frac{g^2 \mu_B^2}{\Delta E}n_\text{imp}$ 
(per magnetic atom in the material).

The quasispins also contribute to the heat capacity of the system. Their contribution is similar to that of an
an ensemble of two-level systems and can be estimated as~\cite{AndersonHalperinVarma}
$C_\text{vac}\sim \frac{T}{\Delta E}n_\text{imp}$ (assuming the non-vanishing density of the effective field acting on the quasispins at small values of the field).
Experimental observation of such a contribution to the heat capacity requires systematically accounting for the large contribution of
bulk spins~\cite{RamirezCooper:SCGOmodes,Nakatsuji:NiGaSmodes,Podolsky:NiGaSmodes,GarattChalker:modes},
as well as the heat capacity of the phonons.


{\it Conclusion.} We have developed a phenomenological theory of the
glass transition and magnetic susceptibility in a geometrically frustrated (GF) magnet
in the presence of vacancies, the most common type of defects in GF magnetic materials.
In our model, the glass transition exists in the absence of vacancies, driven, e.g., by
other sources of quenched disorder. 
Vacancies lead to the formation of quasispin degrees of freedom near the location of the vacancies,
as well as to reducing the density of the spins. The interplay of these effects results in the increase 
of the total magnetic susceptibility of the system.

Increasing a small concentration of vacancies leads to a decrease in the glass-transition temperature, in contrast
with quenched disorder that creates weak local perturbations (e.g. weak random strain).
The described dependencies of the susceptibility and the transition
temperature on the vacancy concentration is consistent with the experimentally observed trends in all GF magnetic materials~\cite{Syzranov:HiddenEnergy}.
The model proposed here and other predictions can be tested in experiment by varying non-vacancy quenched disorder, e.g., by substituting
non-magnetic atoms
in GF magnets by non-magnetic impurity atoms. 
 
We are indebted to A.~Alexandradinata and especially A.P.~Ramirez for useful discussions and comments on the manuscript. 
We acknowledge support by the NSF grant DMR-2218130.

\onecolumngrid
\vspace{2cm}

\cleardoublepage

\setcounter{page}{1}
\pagestyle{empty}

\renewcommand{\theequation}{S\arabic{equation}}
\renewcommand{\thefigure}{S\arabic{figure}}
\renewcommand{\thetable}{S\arabic{table}}
\renewcommand{\thetable}{S\arabic{table}}
\renewcommand{\bibnumfmt}[1]{[S#1]}
\renewcommand{\citenumfont}[1]{S#1}

\setcounter{equation}{0}
\setcounter{figure}{0}
\setcounter{enumiv}{0}

\begin{center}
	\textbf{\large Supplemental Material for \\
		``Effect of vacancy defects on geometrically frustrated magnets''
	}
	\\
	Sergey Syzranov
\end{center}



Here, we provide a generic expression for the correlator $\tilde{K}_{\br\br^\prime}=\langle Q_{\br}^{\alpha\beta}Q_{\br^\prime}^{\alpha\beta}\rangle/T$
of a Gaussian field $Q_{\br}^{\alpha\beta}$ in the presence of multiple randomly located impurities in terms of the correlator ${K}_{\br\br^\prime}$ in an impurity-free systems. The $i$-th impurity 
creates a delta potential at its location $\brho_i$ and adds an energy of $\lambda \left(Q^{\alpha\beta}_{\brho_i}\right)^2$ to the energy 
of the system.
\begin{figure}[H]
	\centering
	\includegraphics[width=0.7\linewidth]{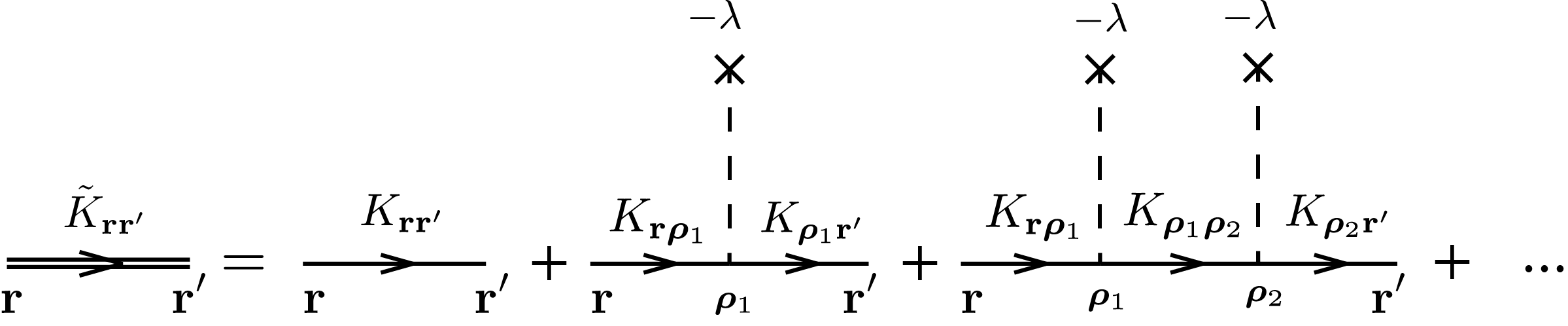}
	\caption{\label{fig:fullk} Perturbative contributions to the correlator
	$\tilde{K}_{\br\br^\prime}$ of the Gaussian field $Q_{\br}^{\alpha\beta}$
	in the presence of randomly located impurities with delta potentials. The impurity locations $\brho_1$, $\brho_2$, $\ldots$
	may coincide.}
\end{figure}

The Gaussian statistics of the field $Q_{\br}^{\alpha\beta}$ allow to compute their Green's function diagrammatically.
The Green's function $\tilde{K}_{\br\br^\prime}$ can be represented as a sum of diagrams in Fig.~\ref{fig:fullk} and is given by  
\begin{align}
	\tilde{K}_{\br\br^\prime}=K_{\br\br^\prime}
	-\lambda\sum_{i,j=1}^N K_{\br\brho_i}\left(1+\lambda\hat K\right)^{-1}_{ij}K_{\brho_j \br^\prime},
\end{align}
where $N$ is the number of the impurities, $\hat K$ is an $N\times N$ matrix with the matrix elements $\left(\hat K\right)_{ij}=K\left(\brho_i-\brho_j\right)$.

If the impurities are dilute, the off-diagonal elements $K(\brho_i-\brho_j)$ of the matrix $\hat K$ can be neglected in comparison with the 
diagonal elements $K(0)$, due to the spatial decay of the correlations between the fields $Q^{\alpha\beta}_\br$. In this dilute limit, the correlator
in the presence of impurities is given by
\begin{align}
	\tilde{K}_{\br\br^\prime}=K_{\br\br^\prime}-\sum_i K_{\br\brho_i}\frac{\lambda}{1+\lambda K_{\brho_i\brho_i}}K_{\brho_i\br^\prime}.
\end{align}

\end{document}